\documentclass{iau} 
\usepackage{graphicx}
 \usepackage[section]{placeins}
 
\title[Maximum possible luminosity of pULXs] %% give here short title %%
{Importance of electron-positron pairs on the maximum possible luminosity of the accretion columns in ULXs}

\author[V. F. Suleimanov, A. Mushtukov, I. Ognev, V. A. Doroshenko \& K. Werner]   %% give here short author list %%
{Valery F. Suleimanov$^{1,2,3}$, Alexander Mushtukov$^{4,3}$, Igor Ognev$^5$,
 Victor A. Doroshenko$^{1,3}$
 \and Klaus Werner$^1$}

\affiliation{$^1$Institut f\"ur Astronomie und Astrophysik, Universitat Tuebingen, \\ Sand 1,
D-72076, T\"ubingen, Germany \\ email: {\tt suleimanov@astro.uni-tuebingen.de} \\[\affilskip]
$^2$Kazan Federal University,  Kremlevskaya 18,
420008 Kazan, Russia \\[\affilskip]
$^3$Space Research Institute of the Russian Academy of Science, \\ Profsoyuznaya 84/32,
117997  Moscow, Russia\\[\affilskip]
$^4$Leiden Observatory, Leiden University, NL-2300RA Leiden, The Netherlands\\[\affilskip]
$^5$P. G. Demidov Yaroslavl State University, Sovietskaya 14, 150003 Yaroslavl, Russia}
\pubyear{2022}
\volume{363}  %% insert here IAU Symposium No.
\jname{Neutron Star Astrophysics at the Crossroads:
Magnetars and the Multimessenger Revolution, IAU Symposium N 363}
\editors{E. Troja \&  M. Baring, eds.}
\begin{document}

\maketitle

\begin{abstract}
One of the models explaining the high luminosity of pulsing ultra-luminous X-ray sources (pULXs) was suggested by
 \cite[Mushtukov et al. (2015)] {Mushtukovetal.15}. 
They showed that the accretion columns on the surfaces of highly magnetized neutron stars can be very luminous due to 
opacity reduction in the high magnetic field.  However, a strong magnetic field leads also to amplification of the 
electron-positron pairs creation. Therefore, increasing of the electron and positron number densities compensates 
the cross-section reduction, and the electron scattering opacity does not decrease with the magnetic field magnification. 
As a result, the maximum possible luminosity of the accretion column does not increase with the magnetic field. It ranges 
between 10$^{40} - 10^{41}$\,erg s$^{-1}$ depending only slightly on the magnetic field strength. 
\end{abstract}
              
\underline{\it Introduction} At present a few pULXs are known (see e.g. Fabrika et al. 2021). One of the models describing 
their observed  high luminosities is a strongly magnetized neutron star (NS)  with high mass-accretion rate 
(Mushtukov et al. 2015) in a binary system. This model predicts increasing of the pULX luminosity with the  strength of 
magnetic field, and, therefore, NSs with a magnetar-like magnetic fields are necessary for explaining the observed pULX 
luminosities (see also Eksi et al. 2015, Brice et al. 2021). Here we improve the model by changing of some geometry 
assumptions and considering the contribution of electron-positron pairs to the opacity.

\underline{\it Model} 
We made following improvements in the cross-section model of the accretion column.
 The  magnetospheric radius   $R_{\rm m}$  in radiation-dominated discs as well as the propeller conditions 
were computed according to Chashkina et al. (2019). We fixed the relative thickness of the transition layer between the disc 
and the magnetosphere $z_{\rm d} = \Delta R / R_{\rm m}$. 
We took into account  the radiation friction between the radiation and  the plasma outside the column. 
We changed the velocity law along the column height, $V(h) = V_0(h/H_{\rm x})^{\xi}$, where $\xi >0$ is a parameter, 
and $H_{\rm x}$ is the current column height, $H_{\rm x=0} = H_0$. The dependence of the 
horizontal radiation flux was taken as $F(x) = F_0(1-\tau_{\rm x}/\tau_0)^{1/\beta}$ with a parameter $\beta \le 1$.
 e$^+$ - e$^-$ pairs were considered  assuming thermodynamic equilibrium according to Kaminker \& Yakovlev (1993) 
 and Mushtukov et al. (2019).   It was shown in these  papers that a number density of the pairs strongly depends 
 on the magnetic field strength and increases significantly if $B > B_{\rm cr}= 4.414\times 10^{13}$\,G. 
 We assume the solar H/He mix as a chemical composition of the accreting plasma, and
 take into account all the opacity sources namely free-free transitions, electron scattering, cyclotron absorption, opacity 
 due to two-photon annihilation and one-photon annihilation in a strong magnetic field.

\begin{figure}[h]
\begin{center}
 \includegraphics[width=2.6in]{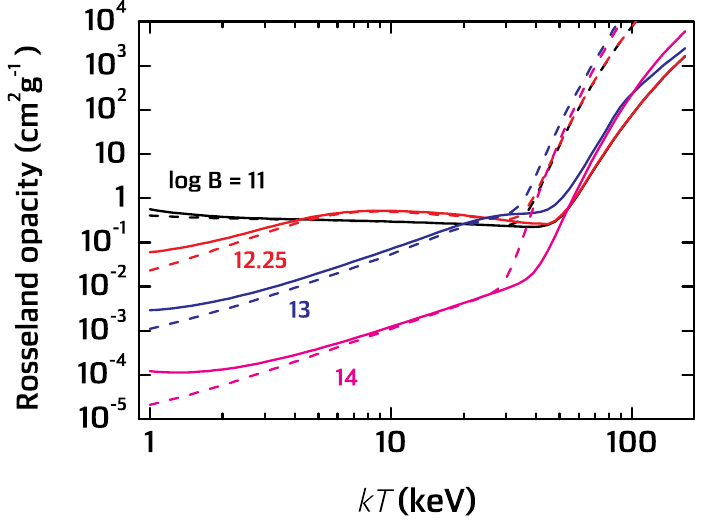} 
\includegraphics[width=2.6in]{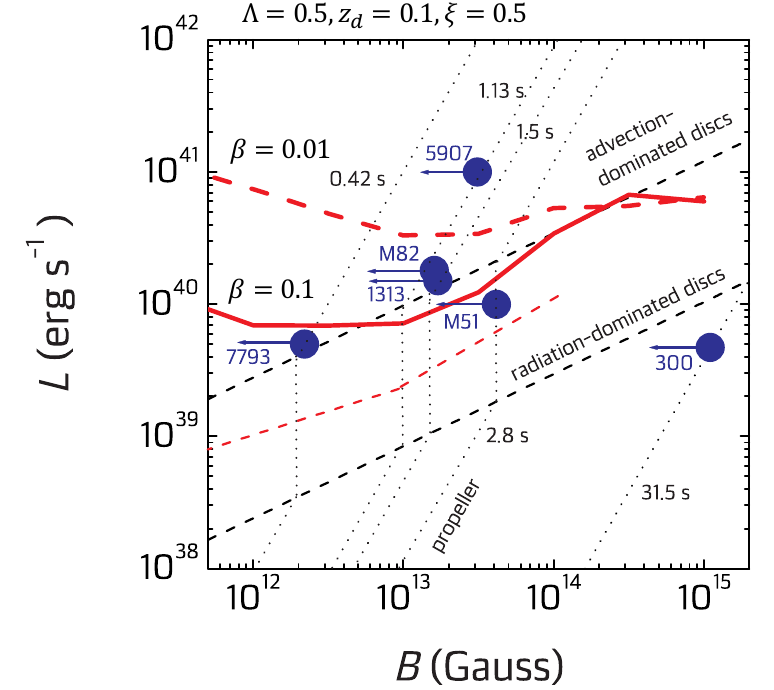} 
\end{center}
 \caption{{\it Left panel:} Dependences of the Rosseland mean opacities on 
the plasma temperature for various magnetic field strengths and two plasma densities, $\rho$= 0.1 (dashed curves) and 10\,g\,cm$^{-3}$ (solid curves). The magnetic field strengths are marked at the curves. 
  {\it Right panel:} Positions of the pULXs on the luminosity - NS magnetic field strength plane. The upper limits on the magnetic
   field strength are found from the propeller conditions. The corresponding conditions for five spin periods are shown with the
    dotted lines. The approximate boundaries for the magnetospheres located in the radiation- and the advection-dominated 
    parts of the discs are also plotted with the dashed and the dot-dashed lines respectively. The new recomputed maximum 
    possible luminosities of the accretion columns (with $H_0=R_{\rm NS}$) are plotted with the thick solid ($\beta$=0.1) 
    and dashed ($\beta$=0.01) red curves. The corresponding old curve (Mushtukov et al. 2015) is drawn with the thin 
    dashed red curve.
   } 
   \label{max_lum}
\end{figure}

\begin{figure}[h]
\begin{center}
 \includegraphics[width=7.2in]{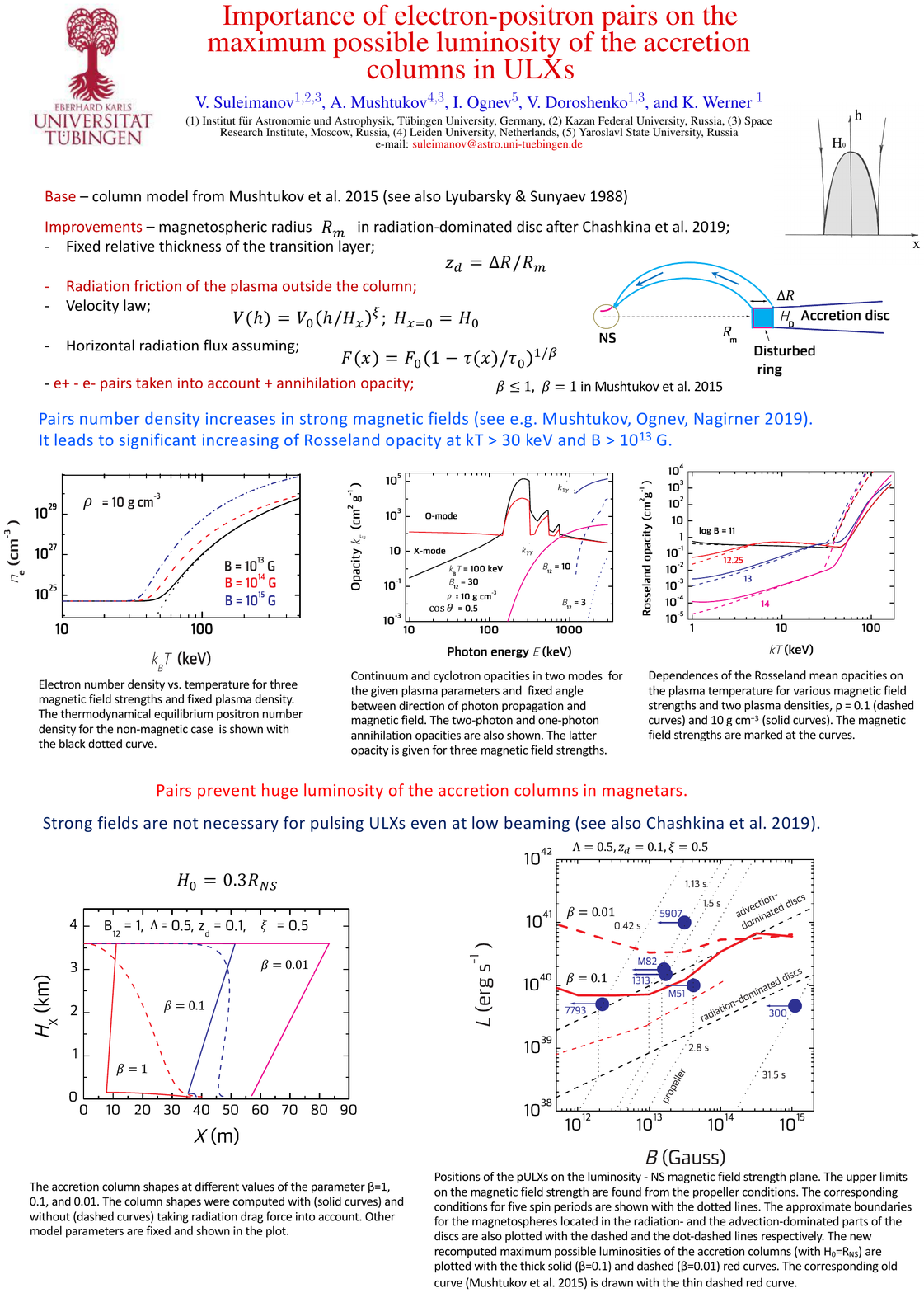} 
\end{center}
 \caption{The poster presented in the  IAU Symposium N 363.  } 
   \label{poster}
\end{figure}

\underline{\it Results} Electron scattering on e$^+$ - e$^-$ pairs increases
the Rosseland opacity at $kT > 30$\,keV significantly (right panel in Fig.\,\ref{max_lum}).
We note that high luminosity accretion columns are optically thick at the considered 
parameters. Therefore, there are conditions for pair thermodynamical equilibrium and the pair numerical densities are
significant along all the column height at high magnetic fields $B > B_{\rm cr}$.
The opacity increase due to electron scattering on the pairs overcomes the opacity reduction due to
magnetic field increase, and pairs prevent a huge luminosity of the accretion columns in magnetars 
 at $B > 10^{14}$\,G. (left panel in Fig.\,\ref{max_lum}).  The considered model provides the high enough 
accretion column luminosities at some parameters even at low magnetic fields, and   
strong NS fields are not necessary for pULXs even at low beaming due to the new propeller conditions derived by
 Chashkina et al. (2019). 
\underline{\it Appendix} The poster itself presented as Fig.\,\ref{poster}.

\end{document}